\documentclass[aps,twocolumn,prb,superscriptaddress,showpacs,eqsecnum]{revtex4}
\usepackage{amsmath,amssymb,latexsym}
\usepackage[dvips]{graphicx}
\usepackage{epsfig}

\begin{document}

\title{Excitons in boron nitride nanotubes: dimensionality effects}

\author{Ludger Wirtz}
\affiliation{Institute for Electronics, Microelectronics, and Nanotechnology
(IEMN), CNRS-UMR 8520, B.P. 60069, 59652 Villeneuve d'Ascq Cedex, France
}

\author{Andrea Marini}
\affiliation{Istituto Nazionale per la Fisica della Materia e 
Dipartimento di Fisica dell'Universit\'a di Roma ``Tor Vergata'',
Via della Ricerca Scientifica, I-00133 Roma, Italy}

\author{Angel Rubio}
\affiliation{Donostia International Physics
Center (DIPC), 20018 Donostia-San Sebasti\'an, Spain}
\affiliation{Department of Material Physics, UPV/EHU and Centro Mixto CSIC-UPV,
20018 San Sebasti\'an, Spain}
\affiliation{Institut f\"ur Theoretische Physik, Freie Universit\"at Berlin,
Arnimallee 14, D-14195 Berlin, Germany}

\date{\today}

\begin{abstract} 
We show that the optical absorption spectra of boron nitride (BN) nanotubes are
dominated by strongly bound excitons. Our first-principles calculations
indicate that the binding energy  for the first and dominant excitonic peak
depends sensitively on the dimensionality of the system, 
varying from $0.7$\,eV in bulk hexagonal BN via 2.1\,eV in the single sheet 
of BN to more than $3$\,eV in the hypothetical $(2,2)$ tube.
The strongly localized nature of this exciton dictates the fast 
convergence of its binding energy with increasing tube
diameter towards the sheet value. 
The absolute position of the first excitonic peak is almost
independent of the tube radius and system dimensionality. This provides an
explanation for the observed ``optical gap'' constancy for different tubes and 
bulk hBN [R. Arenal et al., to appear in Phys. Rev. Lett. (2005)].
\end{abstract}


\maketitle
In complete analogy to carbon nanotubes, boron nitride 
nanotubes \cite{rub94,cho95} can be thought of as cylinders
that are obtained when a single sheet of hexagonal BN is rolled onto itself.
Hexagonal BN (hBN) is a large band-gap insulator\cite{blasebn}. Since
the band-structure of the tubes can be constructed from the
band-structure of the sheet through the zone-folding procedure
\cite{dressbook}, the band-gap of BN tubes is similarly large, 
independently of their radius and chirality.
The detailed knowledge of the optical properties of BN
tubes is indispensable for their characterization and may help to guide 
their use as nanoelectronic devices. E.g., BN nanotubes have been
used to build a field effect transistor \cite{rado}. Furthermore,
experiments on ultraviolet luminescence \cite{wata}
of bulk BN suggest to explore the use of BN nanotubes
as ultraviolet light sources. In this context, it is crucial
to know about possible excitonic states
whose importance has been recently 
shown for the optical spectra of carbon nanotubes \cite{spataru,chang}.
For the wide band-gap BN tubes, we expect even stronger
excitonic effects.

Very recently, two experimental studies
of the optical properties of BN nanotubes have appeared
in this journal which strongly contradict each other. Both studies
compare their spectra to the one of bulk BN which has its first
absorption peak at 6.1 eV and an onset of absorption at about
5.8 eV. Lauret et al. \cite{lauret} have measured two additional
peaks in the optical absorption spectra of BN tubes at 4.45 and 5.5 eV.
The lower of these two peaks was interpreted as a due to a bound
exciton. Arenal et al. \cite{arenal}, on the contrary, have measured the 
electron-energy loss spectra (EELS) of isolated BN tubes and obtained a 
constant ``optical gap'' of 5.8 eV for bulk BN and different single and 
multi-wall tubes. For a proper interpretation of the spectra, one has 
to take into account that already the absorption peak of bulk
hBN at 6.1 eV is due to a strongly bound Frenkel
exciton \cite{arnaud,wirtz}.
The question to be asked is therefore: how does the binding energy
of this exciton change as we compare the quasi-two dimensional
BN sheet and the quasi-1D BN nanotubes with the 3D
bulk BN. Furthermore: Up to which diameter do tubes
exhibit one-dimensional excitonic effects? 
We show in this letter that the excitonic
binding energy increases strongly with lower dimensionality.
At the same time, however, the quasi-particle gap strongly increases
such that the absolute position of the first (excitonic) absorption peak
remains almost constant in agreement with the experiments of
Ref.~\onlinecite{arenal}.
Furthermore, we address the role of dark singlet and
triplet exciton for ultraviolet luminescence.

So far, the optical properties of 
BN nanotubes have only been calculated \cite{mar03,guo} on the level of the
random-phase approximation (RPA), i.e., in the picture of 
independent-particle excitations. 
In this paper we use the Green's function approach~\cite{rmp} of many-body 
perturbation theory to include electron-electron
and electron-hole effects.
Our calculations of the optical absorption spectra, proceed in three steps. 
We first calculate the Kohn-Sham wave-functions of the 
valence band states and a large number of conduction band states using
density functional theory (DFT) in the local-density approximation (LDA)
\cite{kohnsham} using a plane-wave pseudopotential implementation\cite{abinit,pseudo}. 
Within the GW-approximation \cite{rmp}, we then
calculate the quasi-particle energies (``true'' single-particle excitation
energies)\cite{self}.  In the third step,
electron-hole attraction (excitonic effects) is included by
solving the Bethe-Salpeter (BS) equation\cite{rmp}.

Calculation details: 
We use a trigonal array of tubes with minimum inter-wall distance
of 20 a.u. in order to minimize inter-tube interaction and to
simulate as closely as possible the properties of isolated tubes.
The tubes are geometry-optimized (forces on the atoms less than 
$5\times10^{-5}$ a.u.). 
In the GW calculation~\cite{self} we perform a 
``semi-self consistent'' (GW$_0$) calculation by
updating the quasi-particle energies in $G$ (but not in $W$)
until the resulting quasi-particle energies are converged\cite{comment}.
For the optical absorption spectra with polarization along
the tube axis, transitions between the highest $2n$ valence bands 
(the $\pi$ bands) and the lowest $2n$ conduction bands (the $\pi^*$ bands)
are taken into account (the other transitions being forbidden due
to selection rules). 

\begin{figure}
 \epsfig{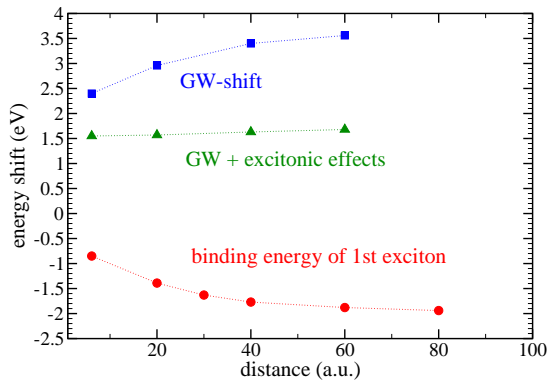}
  \caption{Single sheet of hBN: dependence of the excitonic binding
energy and the quasi-particle gap on the inter-sheet distance in a
supercell-geometry.}
\label{distance}
\end{figure}
\begin{figure*}
\begin{center}
\epsfig{figure=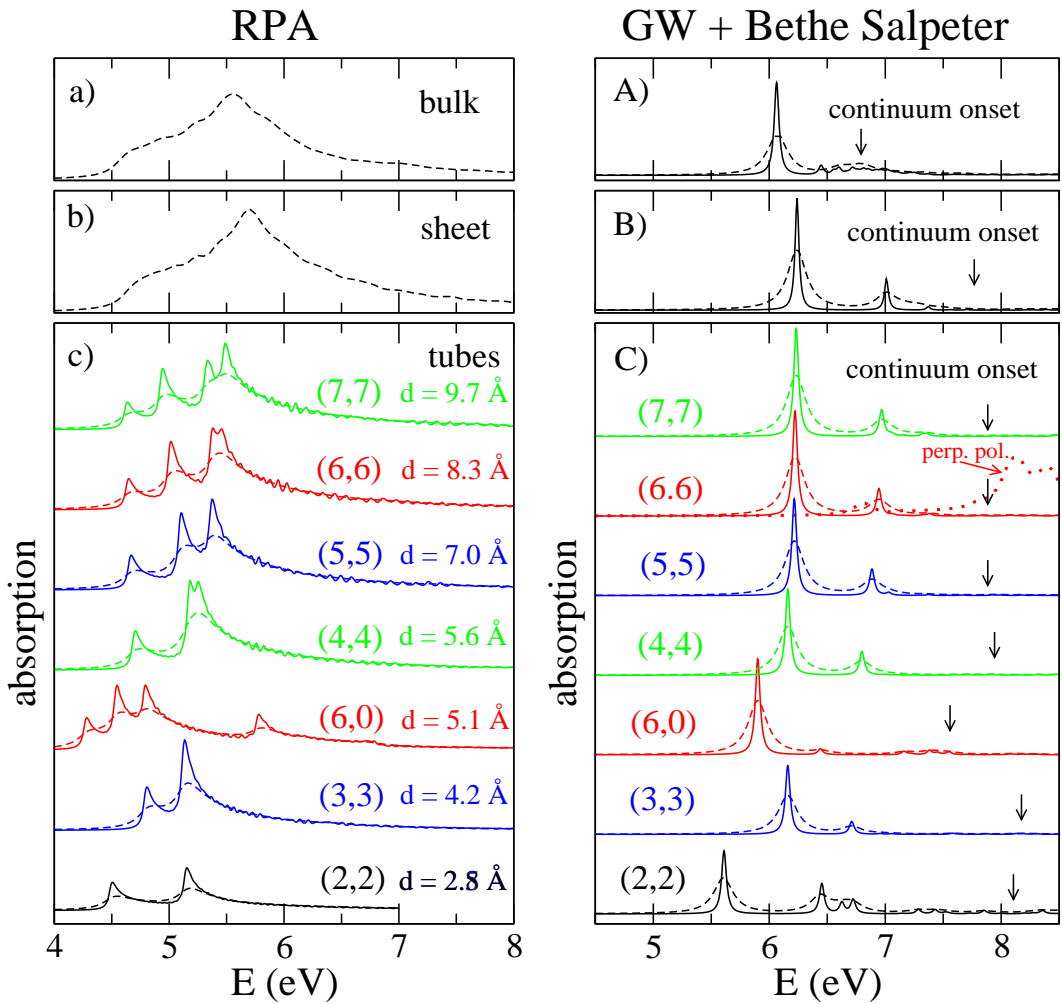,bbllx=32,bblly=106,bburx=573,bbury=684,width=12cm,angle=-90}
\end{center}
\caption{Optical absorption of (a) hBN, (b) BN-sheet, and (c) six different
BN tubes with increasing diameter $d$. We compare the results of the 
GW + Bethe-Salpeter approach
(right hand side) with the random phase approximation (left hand side).
Solid lines are calculated with a Lorentzian broadening of 0.025 eV,
dashed lines with a broadening of 0.1 eV
(for comparison with experimental data).
The light polarization is parallel to the plane/tube axis, respectively
(except for the dotted line in the $(6,6)$ case where the
light polarization is perpendicular to the tube axis, see main text for discussion).
}
\label{spectra}
\end{figure*}

In Fig.~\ref{distance}, we investigate the influence of the
supercell geometry on the excitonic binding energy and on the
quasi-particle gap of the single sheet of hBN. The spectrum
is dominated by the lowest bound exciton which collects most of
the oscillator strength in the energy range between 0 and 8 eV (see
Fig.~\ref{spectra} B). The excitonic binding energy is measured as the
distance between this peak and the onset of the continuum which
is given by the direct quasi-particle gap between the $\pi$ and $\pi^*$
bands. With increasing inter-sheet distance, approaching the limit
of a quasi 2D isolated sheet, the excitonic 
binding energy increases and converges towards the value of 2.1 eV
(as compared to the binding energy of 0.7 eV that is found for the 
3D bulk hBN \cite{wirtz}). 
This increase of the binding energy is due
to two effects: reduced screening for higher inter-sheet distance
and - more importantly - an increased electron-hole overlap in the 
reduced dimensionality (in the purely 2D limit, the binding energy 
for a hydrogenic system is increased by a factor of four compared
to the 3D case \cite{shinada}).
At the same time, the reduced dimensionality leads to an increased
electron-electron correlation and thereby to an increase
of the quasi-particle gap \cite{delerue}. 
Fig.~\ref{distance} demonstrates that the increase of the quasi-particle 
gap almost exactly cancels the increase of the binding energy.
{\it The position of the first absorption peak remains almost
constant}. What changes is the onset of the continuum.
For the BN-sheet however, the absorption at the onset of the continuum
is almost zero (also the higher excitonic peaks carry very low oscillator
strength). The excitonic spectrum can therefore be calculated
to a good approximation already with an inter-sheet distance of 20 a.u.
We made a similar series of calculations for the hypothetical BN(2,2) tube 
which has a diameter of 2.8 {\AA} and is close to being a 1D system.
Again, as we increase the inter-tube distance, the increase of the 
quasi-particle gap almost cancels the increase of the excitonic binding 
energy. While the latter converges towards a value higher than 3\,eV,
the absolute position of the first absorption peak remains constant to within
0.2 eV. In the following, we present therefore calculations for different
tubes in a supercell geometry with 20 {\AA} inter-wall distance.
We remark that dimensionality effects would be more visible 
in other spectroscopic measurements such as photoemission spectroscopy, where
we mainly map the quasiparticle spectra, and this (as the exciton binding
itself) is sensitive to the change in screening going from the tube
to the sheet to bulk hBN. 
In particular the {\it quasi-particle band-gap} will vary strongly with 
dimensionality (opening as dimensionality reduces).

In Fig.~\ref{spectra} we present the spectra of 
bulk hBN, of the single-sheet of hBN and of different BN nanotubes
with diameters ranging from 2.8 {\AA} (for the purely hypothetical
BN(2,2) tube) to 9.7 {\AA} (for the BN(7,7) tube) which is at the lower
border of the range of experimentally produced tubes.
The light polarization is set parallel to the planes or tube-axis, respectively.
On the left hand side, we show the RPA spectra which are almost
indistinguishable for the bulk and for the single-sheet.
The selection rules only allow transitions between the $\pi$ 
and $\pi^*$ bands (band 4 and 5 in the sheet).
The band-structure of the tubes can be constructed via
the zone-folding procedure, i.e., by cutting
the 2D bandstructure of the sheet along certain discrete lines
that correspond to quantized wave-vector along the circumferential
direction. The RPA spectra of the tubes display therefore
transitions at the same energies as in the sheet.
(For a comparison of tube and sheet band-structures,
see Refs.\onlinecite{rub94,blasebn,guo}.)
With increasing diameter, the shape of the tube spectra converges rapidly 
towards the sheet spectrum, in particular if 
plotted with a Lorentzian broadening of 0.1 eV (corresponding
roughly to usual experimental values).
A calculation with a fine
broadening of 0.025 eV (and a correspondingly fine sampling with 200 
${\bf k}$-points in the first Brillouin zone), reveals additional 
fine-structure below 5.5 eV. This structure is due to the van-Hove 
singularities in the one-dimensional density of states. 
For tubes with larger radii, the density of the fine-structure peaks 
increases and the RPA spectrum approaches that of the 2D sheet. 
The onset of absorption is constantly at 4.7$\pm$0.1 eV for all tubes
except for the (2,2) and the (6,0) tube 
(and other small diameter zigzag tubes) where 
the gap is lowered due to curvature effects \cite{rub94}.

While the RPA spectra are due to a continuum of inter-band transitions,
the BS+GW optical spectra on the right-hand side of Fig.~\ref{spectra}
are dominated by discrete excitonic peaks where the first peak
comprises most of the oscillator strength. For bulk hBN, we have
shown \cite{wirtz} that the broadened excitonic spectrum properly
reproduces the experimental spectral shape \cite{tarrio}.
The sheet spectrum contains three bound excitonic peaks of 
rapidly decreasing intensity and absorption at the onset of the
continuum is reduced to almost zero. As explained above, the
stronger binding energy of the first bound exciton is almost
compensated by an increase of the quasi-particle gap due to the
reduced dimensionality. The same holds for the tube spectra:
Except for the three smallest tubes, the position of the first and
dominant excitonic peak remains constant. With increasing tube
diameter, the spectrum rapidly converges towards the three-peak
spectrum of the flat sheet. Also the onset of the continuum 
converges towards the value in the sheet (note that we compare
here the values for a super-cell geometry with inter-sheet/inter-tube
distance of 20 a.u.). The rapid convergence of the excitonic peaks
is an indication for a strong confinement of the exciton wavefunction.
Plotting the wavefunction, we have verified that this Frenkel type 
exciton is confined to within a few inter-atomic distances for 
either tubes, sheet or bulk hBN
(see also the plot for an exciton in bulk hBN in Ref.~\onlinecite{arnaud}).
With increasing tube diameter, the excitons 
only ``see'' a {\it locally flat environment}
which explains the rapid convergence towards the sheet spectrum. 
The strongly localized nature of the
exciton in BN structures makes the appearance of one-dimensional
confinement effects very restricted to small diameter tubes, i.e, tubes
for which the extension of the excitonic wavefunction is comparable to the
nanotube circumference. This is usually the case for carbon nanotubes but
not for the BN tubes as the excitonic wavefunctions only spans a few lattice
constants. As the experimental tubes have diameters around 1.4\,nm, 
the 1D-nature of the tubes cannot be observed and only the 2D nature
of the local exciton environment (tube surface) controls the optical
activity. 

For the (6,6) tube, we display in Fig.~\ref{spectra} C) also
the spectrum for light polarization perpendicular to the tube axis.  
The spectrum exhibits a major excitonic peak that 
lies slightly below the second excitonic peak obtained for light
polarized along the tube axis. 
Note that the sheet is completely transparent up to 9 eV
for light polarized perpendicular to the plane.

We compare our results now to two recent contradictory measurements
of the optical properties of BN nanotubes \cite{arenal,lauret}:
In the EELS experiment of Arenal et al.\cite{arenal},
the electron-beam passes the tube in the tangential direction. 
A detailed explanation of the spectral shape 
would require the calculation of the imaginary part of the
polarizability $\alpha_{m,k}$, where $m$ is the index for the
multi-pole expansion in circumferential direction and $k$ is
the Fourier expansion along the tube axis \cite{taverna}.
This is beyond the scope of this paper, where we only calculate
the dipolar contribution ($m = 0$) in the limit $k \rightarrow 0$.
This is, however, the dominant part in the expansion of $\alpha$.
The constancy of the first excitonic peak explains why the
``optical gap'' observed in Ref.~\onlinecite{arenal}
is always 5.8 eV, whether they measure bulk hBN, multi-wall tubes, 
or single-wall tubes.
We note that there is a chirality dependence of the optical spectra
but it is only visible for the smallest diameter tubes.
The spectra of the armchair tubes converge much faster to the 2D case
than the ones of the zig-zag tubes. However, as experimental tube 
diameters \cite{arenal} are much larger than the 5.1 {\AA} of
the (6,0) zigzag tube that is presented in this paper,
we expect the chiral dependence to be marginal.
Our calculations show that the explanation of Ref.~\onlinecite{lauret}
for the two peaks at 4.45 and 5.5 eV in their absorption spectra
of a sample containing BN tubes does not hold: the peaks are neither
due to additional Van-Hove singularities (since the spectra are
entirely dominated by discrete excitonic peaks) nor can they
be explained by an increased excitonic binding-energy (which
is canceled by an increased quasi-particle gap).

So far, we have concentrated on singlet active excitons. 
For bulk hBN \cite{wirtz}, we have shown previously that there is
a dark singlet exciton and two triplet excitons below
the first optically active exciton.
For the single sheet (and light polarization parallel to the plane), 
we find that the lowest optically active exciton is doubly degenerate.
There is no dark singlet exciton below, but
a doubly degenerate triplet exciton at 0.1\,eV lower energy.
For the (6,6) tube (and light polarization parallel to the tube
axis), we find that the degeneracy of the singlet exciton
is lifted, leading to a dark singlet exciton slightly (0.01\,eV)
below the optically active singlet exciton. The degeneracy
of the triplet exciton is lifted as well: the two triplet
excitons are 0.1\,eV and 0.08\,eV lower in energy than the
optically active singlet exciton. Similar results hold
for the (5,5) and the (7,7) tubes.
A recent study for C-tubes\cite{louie} has shown that the room temperature 
luminescence is enhanced once the complete series of active and dark excitons 
is taken into account. This would hold also in the present case. Furthermore,
due to the minor differences in the optical 
spectra of tubes and bulk BN we expect the BN-tubes to exhibit
a strong ultraviolet lasing behavior as already observed for bulk BN\cite{wata}.
The fact that this luminescence response would be rather insensitive
to tube diameter and chirality makes the BN tubes ideal candidates for
optical devices in the UV regime as the carbon nanotubes are in the infrared
regime\cite{avouris}. The photoluminescence quantum yield of BN tubes
should surpass the efficiency of carbon\cite{wata}.

In conclusion, the optical absorption spectra of hBN and BN
nanotubes are dominated by excitonic effects. Most oscillator strength
is collected by the first bound exciton. Its binding energy 
increases strongly as the dimensionality is reduced from the
3-D bulk over the 2-D sheet to the 1-D tubes. At the
same time the quasi-particle band gap increases with reduced
dimensionality. This cancellation leaves the absolute position
of the dominant absorption peak almost constant. 

We acknowledge helpful discussions with D. Varsano, V. Olevano,
and L. Reining.
The work was supported by the EU network of excellence 
NANOQUANTA (NMP4-CT-2004-500198) and the French GDR "nanotubes".
Calculations were performed at IDRIS (Project No. 51827) and CEPBA.
A.R. acknowledges the
Humboldt Foundation under the Bessel research award (2005),


\begin{thebibliography}{99}
\bibitem{pseudo} We used Trouiller-Martins pseudopotentials
in the Kleinman-Bylander form with cutoff radius of 1.59 a.u. for B
and 1.50 a.u. for N. An energy cutoff of 25 Hartree is used in the 
plane-wave expansion of the wave-functions.
\bibitem{comment}
This procedure yields a quasi-particle bandgap for the single
sheet and for the tubes that is about 0.25 eV higher than the
one obtained on the non-self consistent (G$_0$W$_0$) level.
Still, for bulk hBN we have shown\cite{wirtz} that 
this GW$_0$ calculation starting from LDA wave-functions yields 
a quasi-particle band-gap that is too low by 0.65 eV. 
In the present work, we assume that the same underestimation of the 
LDA GW$_0$ band-gap holds also for the single sheet and for the tubes.
Consequently, we add a 0.65 eV ``scissor'' shift to the calculated
GW$_0$ band-structure.
\end{thebibliography}
\end{document}